# Tunneling spectroscopic signatures of charge doping and associated Mott transition in α-RuCl$_3$ in proximity to graphite

Xiaohu Zheng[1,2*], Ke Jia[3], Junhai Ren[1], Chongli Yang[1], Xingjun Wu[1], Youguo Shi[3], Katsumi Tanigaki[1], Rui-Rui Du[2,4]

[1]Beijing Academy of Quantum Information Sciences, Beijing 100193, China.

[2]International Center for Quantum Materials, School of Physics, Peking University, Beijing 100871, China.

[3]Beijing National Laboratory for Condensed Matter Physics and Institute of Physics, Chinese Academy of Sciences, Beijing 100190, People's Republic of China.

[4]CAS Center for Excellence in Topological Quantum Computation, University of Chinese Academy of Sciences, Beijing 100190, China.

**ABSTRACT: The layered Mott insulator α-RuCl$_3$ has been extensively studied as a potential Kitaev quantum spin liquid candidate. Here, by constructing heterostructures with graphite, we employed electron tunneling measurements on few-layer α-RuCl$_3$ using a scanning tunneling microscopy/spectroscopy. Characteristic tunneling spectra were detected on α-RuCl$_3$ layers in proximity to graphite. In the single-layer α-RuCl$_3$ in direct contact with graphite, distinct states in the Mott-gap regime were observed. The in-gap states are demonstrated to be closely related to the electron orbitals in α-RuCl$_3$ and graphite, and to be sensitive to interfacial coupling, where a hybridization at the heterointerface is hypothesized. The in-gap states are also thought of as a charge reservoir for weakly doping the α-RuCl$_3$ upper-layers. It demonstrated that the weak doping effect causes a considerable decrease in the Mott-gap within the upper-layers, suggesting that an unconventional Mott-** **transition is occurring in these layers. The results show that the heterostructure comprised of α-RuCl$_3$ and graphite is a good platform for investigating the doping physics in α-RuCl$_3$. Therefore, tunneling into such a doped system is a useful probe for studying otherwise insulating spin-liquid candidates.**

**INTRODUCTION:** Kitaev quantum spin liquid (QSL) is a theoretical model of a strongly correlated spin phase. Its exact solution predicts the presence of Majorana zero mode (MZM) that obeys non-Abelian statistics, which could pave the way for fault-tolerant quantum computation [1,2]. A layered Mott insulator, α-RuCl$_3$ has been extensively investigated as a possible candidate for Kitaev QSL [3–8]. The fingerprints for Majorana fermions of the fractionalized spin excitations in α-RuCl$_3$ have been reported by several experimental groups [9–11] Recent measurements of thermal Hall conductance with half-integer quantization, in particular, have provided important evidence for the chiral edge modes of Majorana fermions [12–20]. On the other hand, the roadmap designed for quantum technology platforms typically relies on electronic methods to manipulate the quantum bits [21]. The chargeless character of quasiparticles and the electrically insulating nature of Kitaev QSL materials like α-RuCl$_3$, would limit a range of suitable electronic measuring techniques. Recently, several experiment setups in heterostructures comprised of conducting or superconducting films [22–31] have been proposed to exhibit electronic accessible signs of the quantum excitations in Kitaev QSL. Additionally, the doped Kitaev models are also thought to host a number of exotic quantum phases, such as p-wave superconductivity [32–34], making charge doping in Kitaev QSL research more appealing and essential. On the experimental side, several groups have reported the charge doping of Kitaev materials *via* forming heterostructures with graphene [35–38], or ionic intercalations [39], but the doping mechanism of the Mott-insulating Kitaev honeycomb lattice remains elusive.

In this research, we transferred Kitaev QSL candidate material α-RuCl$_3$ flakes onto a highly-oriented pyrolytic graphite (HOPG) substrate to create heterostructures, as illustrated schematically in Fig. 1a. The surface morphology and electronic local density of states (LDOS) of α-RuCl$_3$ were evaluated by electron tunneling through the thin film using scanning tunneling microscopy and spectroscopy (STM/STS) after annealing the samples at 280 °C in an ultra-high vacuum chamber for 2 hours. We have attempted tunneling experiments on samples thicker than 20 nm; however, STM/STS can only be conducted at room temperature, on those flakes. Thus, all the samples in this study are all below 20 nm. The lock-in technique was used to measure STS at a frequency of 707 Hz and modulation amplitudes ranging from 5 mV to 8 mV. On *dI/dV* spectra, distinct states in the Mott-gap regime were observed in the first (1$^{st}$) α-RuCl$_3$ layer that directly contacts to graphite. It is demonstrated that the in-gap states result from orbital hybridization at the heterointerface. Surprisingly, a few upper layers exhibit a completely different spectral phenomenology than the bulk α-RuCl$_3$ due to weak charge transfer from the heterointerface, where the Mott-gap is significantly reduced. It implies that doping charges and strong quantum-fluctuation in α-RuCl$_3$ combine to form a novel correlated insulating state with a reduced energy gap.

**RESULTS and DISCUSSION:** α-RuCl$_3$ is an insulating 4*d* transition-metal halide with a honeycomb lattice composed of nearly ideal edge-sharing RuCl$_6$ octahedra [40,41]. Its basic structure is similar to that of graphite, where the bulk crystal can be exfoliated into a fully two-dimensional magnet, and the Ru atoms surrounded by Cl atoms form the 2D honeycomb lattice, as shown inset in Fig. 1b. The amplitude and characteristics of the bandgap, however, remain debatable; for example, the gap amplitude has been reported to range from 0.2 to 1.9 eV depending on the methodologies used [7,8,42,43]. In Fig. 1b, an atomic-resolved STM morphology was taken on an α-RuCl$_3$ flake with a thickness more than 5 nm, which shows intrinsic properties of the bulk α-RuCl$_3$. A honeycomb lattice with several vacancies were observed in Fig. 1b. The lattice constant is approximately 6 Å (Fig. 1c), which is consistent with the theoretical prediction [41]. The averaged *dI/dV* spectrum measured at 77 K which is far above the Néel temperature ($T_N$~7 K) [11], reveals a U-shaped bandgap of ~1.8 eV throughout the entire field of view (Fig. 1d). In our experiment, the spectra are fully gapped without any single-particle excitations in the gap regime (Fig. 1d), and scarcely affected by disorders (Figs. e and f), showing a clear divergence from the spectrum taken at room temperature [42]. It coincides with the consensus that bulk α-RuCl$_3$ is a spin-orbital assisted Mott insulator in paramagnet phase [3], with significant Kitaev spin correlations that persist over a temperature range of 7 to 120 K [11,44–46]. Thus, in the spectra in Fig. 1d, conduction (CB) and valence bands (VB) with sharp edges correspond to the upper and lower Hubbard bands (UHB and LHB). The amplitude of the Mott-gap is consistent with optically probed values [7], and calculations using the local density approximation (LDA) [7,47]. Overall, our tunneling experimental results on thick α-RuCl$_3$ flakes measured at 77 K have clearly highlighted the strong correlations and the Mott insulating natures of the bulk α-RuCl$_3$ [8,43].

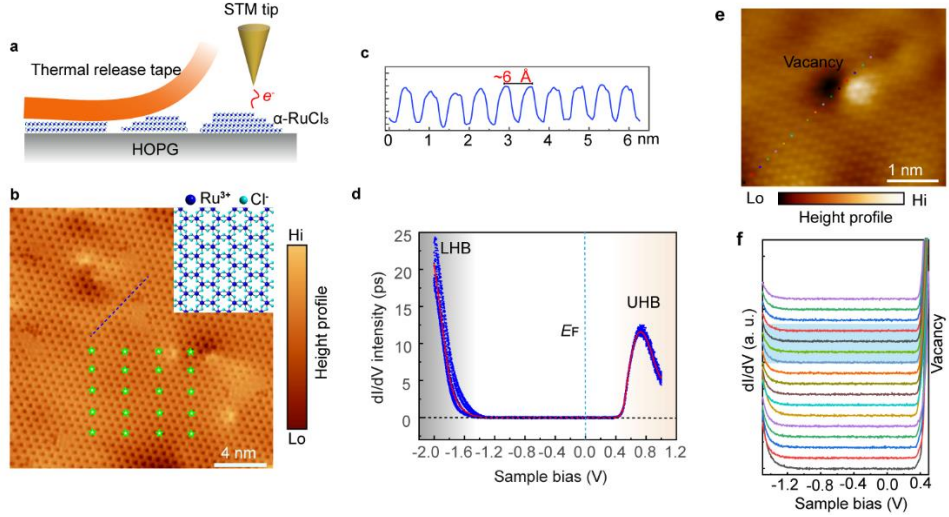

**Figure 1. a**, Schematic diagram of the experimental setup where the α-RuCl$_3$ flakes with different thicknesses are transferred by thermal release tap and measured through electron tunneling by STM/STS; **b,** STM morphology of the bulk α-RuCl$_3$ (thickness > 5 nm), the lattice construction are presented inset (sample voltage: $V_H$=800 mV; setpoint current: $I_s$=0.6 nA); **c,** The cross-sectional profile along the blue line shows the lattice constant is about 6 Å; **d**, *dI/dV* spectra taken at different locations (labeled with stars) on the surface in (b) (dark blue curves) and an averaged *dI/dV* spectrum (red) show a full bandgap of about 1.8 eV±0.2 eV ($V_H$=800 mV, $I_s$=0.6 nA). The colored shadow regions show the Hubbard bands; **e**, STM morphology shows a single-vacancy on α-RuCl$_3$ flake ($V_H$=550 mV, $I_s$=1 nA); **f**, *dI/dV* spectral curves were collected at the locations as labeled by the corresponding colored dots in (e), the spectra taken near the vacancy (shadowed with light blue) show negligible difference without any featured DOS arising in the Mott-gap regime ($V_H$=550 mV, $I_s$=1 nA).

As described in refs. [32–35,48], charge transfer occurs when the heterostructure is made between α-RuCl$_3$ and graphene. The doping mechanism and electronic states of the weakly doped α-RuCl$_3$ are, however, still unclear. Here, in the samples of α-RuCl$_3$ on graphite, we focused on the thin α-RuCl$_3$ flakes that were in proximity to graphite. As shown in Fig. 2a, a boundary region containing α-RuCl$_3$ terraces of single-layer (denoted as 1:α-RuCl$_3$) and bilayer (2:α-RuCl$_3$) on graphite was acquired with a broad-view STM image. The height-profile of single-layer and bilayer α-RuCl$_3$ is approximately 600 pm (upper panel in Fig. 2b), which is consistent with the anticipated thickness of single-layer α-RuCl$_3$ in van der Waals interlayer coupling [5,49]. We note that the thickness of monolayer α-RuCl$_3$ on graphite is only 350 pm, which is significantly less than the predicted value [5,49]. Despite the possibility of errors caused by the huge DOS difference (insulating *vs* metallic) on the neighboring sides, the interfacial coupling between graphite and α-RuCl$_3$ may be stronger than the general van der Waals due to probable interfacial hybridization. As demonstrated in Fig. 2c, atomic-resolved STM images were employed on terraces of 1:α-RuCl$_3$ and 2:α-RuCl$_3$, respectively. Spatial *dI/dV* spectra were collected in labeled positions on 1:α-RuCl$_3$. Despite the fact that all the tunneling curves mimic the Mottness spectra with a large gap, as depicted in the left panel in Fig. 2d, in-gap states were observed in an extended view of the gap regime (inset in Fig. 2d). The in-gap states show spatially inhomogeneous intensities as moving tip positions, as shown inset in Fig. 2d. The *dI/dV* spectra on the terrace of 2:α-RuCl$_3$ have a completely different line-shape than those taken on the thick α-RuCl$_3$ and the single-layer flakes, and are uniform over the entire surface, where the energy-gap is commonly and substantially reduced (~150 meV), as shown in the right panel of Fig. 2d. The tunneling signatures in Fig. 2d indicate that considerable modifications of the electronic states are occurring in the 1$^{st}$ and 2$^{nd}$ α-RuCl$_3$ layers on graphite.

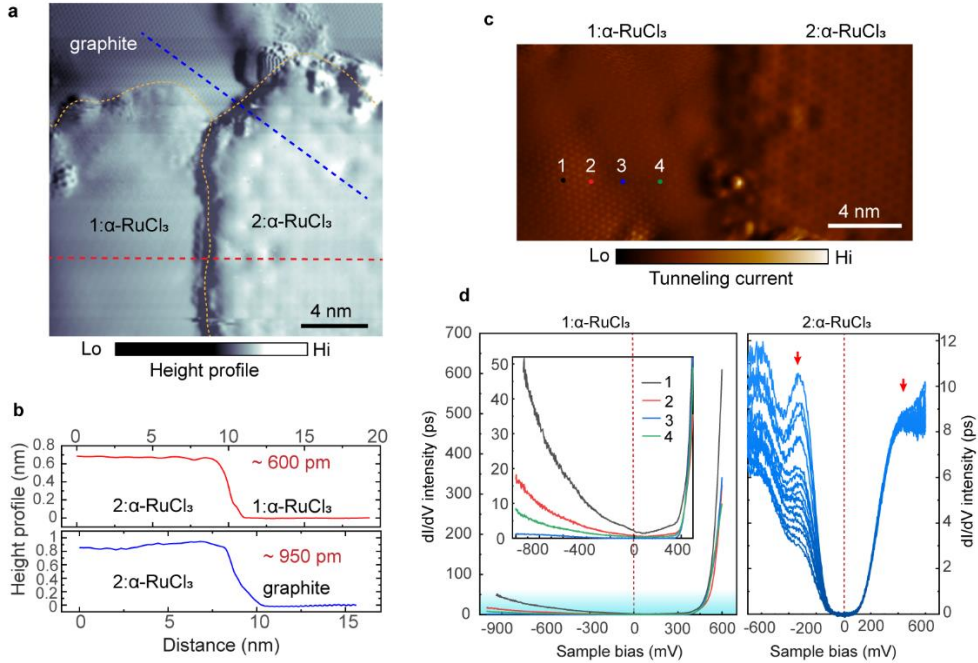

**Figure 2. a**, A broad view STM image of the graphite surface covered by α-RuCl$_3$ flake with single and bilayer terraces (denoted as 1:α-RuCl$_3$ and 2:α-RuCl$_3$) ($V_H$ =800 mV; $I_s$=200 pA); **b**, Step profiles acquired between 2:α-RuCl$_3$ and 1:α-RuCl$_3$, 2:α-RuCl$_3$ and graphite, respectively; **c**, High-resolution STM image across 1:α-RuCl$_3$ and 2:α-RuCl$_3$ ($V_H$ =500 mV; $I_s$=500 pA); **d**, Left: spatial *dI/dV* curves taken at numbered positions in (c) of the 1:α-RuCl$_3$ show the in-gap states are arising and the variations of spectra at different locations, enlarged-view of the shadow regime are shown inset. Right: a series of *dI/dV* spectra taken on the surface of the 2:α-RuCl$_3$ show the uniform reduced-gap spectral curves with a fluctuation of the occupied states ($V_H$=500 mV, $I_s$=0.5 nA).

We postulate that the modifications of α-RuCl$_3$'s electronic structures are caused by the heterostructures with graphite. As a result, the 1$^{st}$ layer that directly contacts graphite is critical for understanding the underlying physics. As shown in the atomic-resolved STM images taken on the 1$^{st}$ α-RuCl$_3$ layer in Fig. 3a, regions with mild lattice distortion are commonly observed due to the inhomogeneous coupling between α-RuCl$_3$ and graphite (artificially created by the transfer process). The appearance of the graphite-like lattice in the distortion zone (Fig. 3a) is fascinating. We carefully monitored a succession of tunneling spectra along the route, as shown in Fig. 3a (Nos. 1 to 10) under fixed tunneling parameters. When the positions are widely apart, the spectra exhibit U-shaped Mott-gap (from No.1 to 4 in Fig. 3b) which are analogous to those on bulk α-RuCl$_3$ (Fig.3b). When the tip reaches the distortion (numbers 5 -10 in Fig. 3b) a broad in-gap state above the LHB arises and extends toward the $E_F$ with a small peak developed at ~-600 meV. There is essentially no change in the number of empty states along the path. The spectral phenomenology described above shows that hole puddles arise at the distortion regions on the surface of the 1$^{st}$ α-RuCl$_3$ layer. It explains the spatially inhomogeneous LDOS seen in Fig. 2d (left).

What about the condition outside the distortion regime if the hole puddles are a consequence of the hybridization at the heterointerface? We discovered that the tunneling spectra are affected by the tunneling parameters, specifically the bias voltage $V_H$. As illustrated in Fig. 3c, two representative sites, labeled No.1 and No. 7, are chosen for measuring the *dI/dV* spectra under different tunneling conditions. Fig. 3c shows dramatically improved in-gap states when $V_H$ decreases from 650 mV to 560 mV in Fig. 3c, where the gapped spectra practically turn to gapless. Such a huge variance in tunneling spectra cannot be interpreted by tip-induced band bending [50,51]. It implies that in-gap states exist even outside of the puddles on the α-RuCl$_3$ single-layer. As a result, we hypothesize that hybridization commonly occurs at the heterointerface.

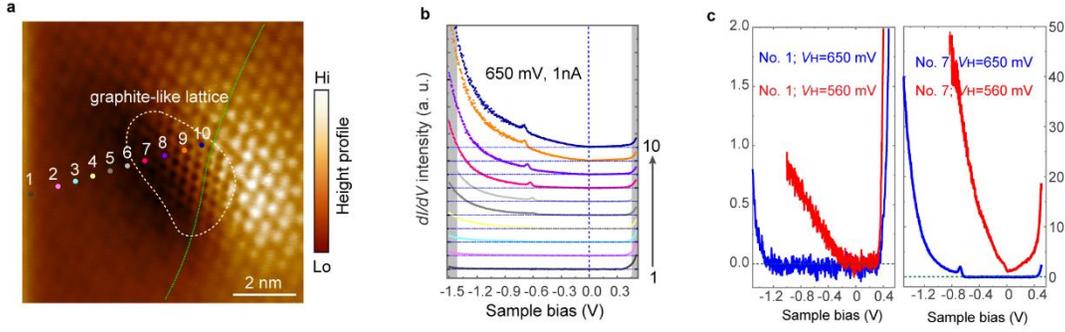

**Figure 3. a**, Atomic-resolved STM image are taken on 1:a-RuCl$_3$ ($V_H$ =600 mV; $I_s$=1 nA), distortion of the lattice along the dashed green line can be seen, and the center of distortion (round with a dashed white line) presents a graphite-like lattice; **b**, The *dI/dV* spectra corresponding to the positions marked by colored dots (No. 1 to No. 10) in (a) show a clear evolution of LDOS. ($V_H$=650 mV, $I_s$=1 nA); **c**, Comparison of *dI/dV* spectra collected at fixed positions No. 1 (left panel) and No. 7 (right panel) as labeled in (a) at different tunneling conditions by changing $V_H$ from 650 mV to 560 mV with a fixed $I_s$=1 nA.

To further comprehend the tunneling signal as the setup parameters ($V_H$) varied, atomic-resolved STM images were taken in a fixed area (Fig. 4a) on 1:a-RuCl$_3$ with varying $V_H$. When $V_H$=800 mV and $I_s$=600 pA, the lattice of α-RuCl$_3$ is clearly visible across the entire region, as shown in Fig. 4b. When $V_H$ is reduced to 600 mV, a tiny distortion regime (DOS puddle) surrounded by α-RuCl$_3$ lattice forms in STM morphology, displaying the graphite-like lattice (Fig. 4c), acting similarly to that shown in Fig. 3a. As $V_H$ is decreased with a fixed $I_s$, the graphite-like lattice continually expands, eventually presenting on the entire area of the in-view (from Figs. 4d to g). It demonstrates the fact that hybridization occurs throughout the heterointerface, but the intensity of the hybridized states is tightly related to the interface coupling. As a result, the in-gap states in the region without distortion are faint and observable in low $V_H$ (small tip-surface distance). We validated that the lattice constant and orientation of the graphite-like lattices are entirely consistent with the underneath graphite, as shown in Fig. 4a, As is generally known, STM morphology is a convolution of the scanned surface's topography and electronic states. When the electronic wave functions (EWFs) from the beneath layer successfully dominate the tunneling current, the underneath topographic signals can be disclosed on top of heterostructures [52]. Even at a low sample voltage ($V_H$=410 mV in Fig. 4g), the associated energy remains outside the Mott-gap on 1:a-RuCl$_3$ (in UHB). It suggests that at the heterointerface, overlapping charge orbitals from both α-RuCl$_3$ and graphite layers contributed to the tunneling signals on the surface, resulting in the graphite-like lattice architecture.

We took the *dI/dV* spectra at the locations indicated by the stars in Fig. 4d with different tunneling parameters to observe the LDOS corresponding to the STM topographic evolution. In Fig. 4h (left), we see that decreasing $V_H$ with a fixed $I_s$ results in a peaked DOS at ~-1350 meV above the LHB at $V_H < 550\ mV$. We started to witness the graphite-like lattice emerging in the selected point in Fig. 4e in an approximate $V_H$, indicating that the distinct in-gap state accounts for the emergence of graphite-like lattice on α-RuCl$_3$ surface. However, by comparing the in-gap states to the DOS in the *dI/dV* spectrum of bare graphite, the hypothesis that the peaked states originate solely from beneath the graphite may be ruled out, as shown in right panel in Fig. 4h. It also implies that the peaked states are caused by the heterointerfacial hybridization. As $V_H$ decreases, the in-gap state arises quickly above LHB and extends to $E_F$. Another minor LDOS peak at -600 meV, shown in Fig. 3b, reappears in Fig. 4h (left) with $V_H$ = 450 mV. As $V_H$ decreases further, the peaked LDOS at -600 meV submerges into the fast increasing in-gap states. At this point, the spectrum is substantially equal to the spectrum shown in the right panel in Fig. 3c (red curve), suggesting that the LDOS outside of the lattice distortion are nearly identical to those in the distortion region within the decreased $V_H$. On the other hand, while the gap can be completely closed with finite state at $E_F$ at $V_H < 450\ mV$, both the UHB and LHB are still conserved in a broadly-viewed energy scale (Fig. 4h), indicating that the hybridization didn't collapse the Mott-Hubbard state.

We performed the following study to get compelling proof that the in-gap states are caused by heterointerfacial hybridization rather than tip-surface interactions. The tunneling parameters $V_H$ and $I_s$ were used to position the STM tip on the surface at a specific distance $d_s$ (inset in Fig. 3i), and the spectra were acquired by sweeping the sample bias ($V_b$) while suspending the feedback loop. $d_s$ can be quantitatively calculated using the formula: $d_s = -\frac{\hbar}{2\sqrt{2m\Phi}}\ln\left(\frac{R_0 I_s}{V_H}\right)$, where $m$ is the mass of the tunneling electron, $\Phi$ is the average work function of the tip and the sample, and $R_0$ is the resistance for a single-atomic point contact [53]. The Tersoff and Hamann's tunneling formalism [54] can thus be used to explain $dI/dV$ spectra associated with $d_s$: $\frac{dI}{dV} \propto e^{-2kd_s} N(eU_{bias})$, where $k$ is a constant, and $N(eU_{bias})$ is the surface LDOS at a certain bias voltage $U_{bia}$. $dI/dV$ spectra would be tunable by modulating the work function ($\Phi$) of the tip or introducing the tip-surface interactions ($R_0$), such as the spreading resistance effect [51] and the tip-induced band bending [50]. Indeed, the tip-surface interactions increase DOS amplitude and shift band edges in the $dI/dV$ spectra on α-RuCl$_3$ flakes with different thicknesses (Figs. 4h and j), but they never introduce distinct characteristics (peaked DOS) to the spectra curves. By making the parameters related to the tip and tip-surface interaction constant, the tip-dominated states can be avoided. As a result, only true surface states can be well fitted using Tersoff and Hamann's formalism [54]. In Fig. 4i, we fit the data using the distinct state that peaked at -1350 meV. The growing tendency of peak amplitude over $V_H$ fits nicely, proving that the in-gap states truly existed. However, as shown schematically in Fig. 4i, they have a short decay length. The results in Fig. 4 demonstrated that EWFs from both layers overlap, resulting in the formation of an occupied subband that peaked at -1350 meV above the LHB.

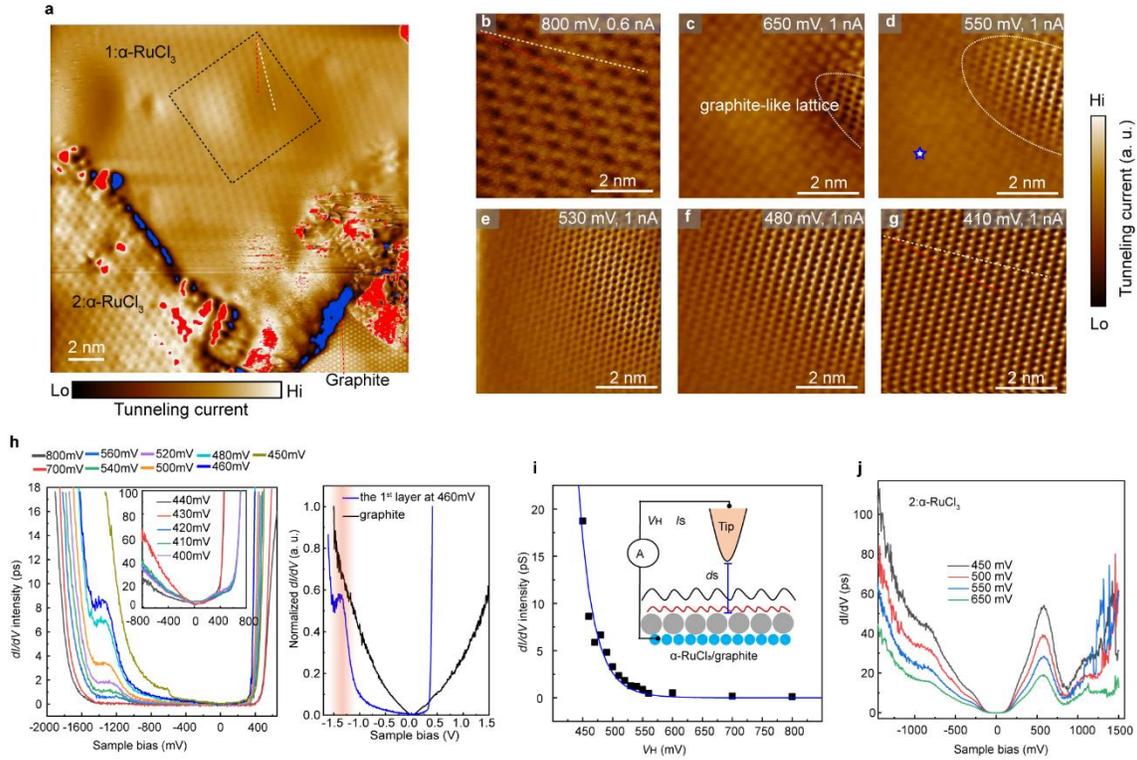

**Figure 4. a**, A broad-view STM image shows lattice topographies of the single-layer (1: α-RuCl$_3$), bilayer α-RuCl$_3$(2: α-RuCl$_3$) and graphene substrate. The lattice orientations in the three regions can be distinguished, where the white and red dashed lines represent the lattice orientations of 1: α-RuCl$_3$ and graphite, respectively; **b-g**, STM morphologies taken on 1:α-RuCl$_3$ with variation of tunneling conditions show evolution from α-RuCl$_3$ to the graphite-like lattice, the evolution begins at the region with slight lattice distortion; **h,** Left: $dI/dV$ spectra on 1:α-RuCl$_3$ acquired with changing $V_H$ from 800 mV to 400 mV with a fixed $I_s$=1 nA, in-gap states peaked at -1350 meV arising in the Mott-gap; Right: normalized $dI/dV$ spectrum taken on the 1$^{st}$ α-RuCl$_3$ layer with $V_H$=460 mV and $I_s$=1 nA, is compared to the normalized spectrum taken on a bare graphite ($V_H$=1V and $I_s$=0.5 nA); **i**, The peak magnitude at -1350 meV as a function of $V_H$ is fitted using the distance-

dependent Tersoff and Hamann's tunneling formalism [54]. Inset schematically shows the experimental setup when $dI/dV$ were performed, where the orbitals can be selectively detected by tunning $V_H$; **j**, $dI/dV$ spectra were taken on the 2:α-RuCl$_3$ with variation of $V_H$ with a fixed $I_s$=1 nA.

An intriguing question arises: may charges from the in-gap states at the heterointerface be transferred to the upper α-RuCl$_3$ layers? Is the doping effect, if so, a precursor to a new electronic phase? The completely distinct $dI/dV$ spectra on the 2$^{nd}$ α-RuCl$_3$ layer (in right of Fig. 2d) appear promising. As a result, the tunneling spectra on few-layer α-RuCl$_3$ must be investigated. Fig. 5a shows a different α-RuCl$_3$ flake on graphite. The surface of the terraces in the square (designated in Fig. 5a) are the 3$^{rd}$ and the 4$^{th}$ α-RuCl$_3$ layers above the graphite (left panel of Fig. 5c), is illustrated by a height profile (Fig. 5b) along the dashed line in Fig. 5a. Unlike the bulk morphology of the bulk α-RuCl$_3$ where the honeycombs are constructed by Ru atoms as seen in Fig. 1b, the STM image on the 3$^{rd}$ layer reveals a lattice structure dominated by the Cl atoms, as shown in the right panel of Fig. 5c. It means that the surface electronic states of the 3$^{rd}$ layer have changed dramatically. As illustrated in Figs. 5d and e, averaged $dI/dV$ spectra are taken on the two terrace surfaces (the 3$^{rd}$ and the 4$^{th}$ layers, respectively). The 4$^{th}$ layer retains a U-shaped spectral curve (> 1.4 eV) similar to the spectra on the bulk α-RuCl$_3$, whereas, the 3$^{rd}$ layer has a spectrum with a suppressed gap (~330 meV) qualitatively similar to the spectra on the 2$^{nd}$ layer in Fig. 1d.

Fig. 5f shows another flake containing surface of the 2$^{nd}$ and 3$^{rd}$ α-RuCl$_3$. Because of the inhomogeneous interface coupling in different flakes, the surface of the 2$^{nd}$ layer has a thickness of ~1.1 nm (Fig. 5g), which is slightly thicker than that in Fig. 2b. The reduced-gap spectra were observed on the 2$^{nd}$ layer (Fig. 5h) in this region, confirming a common characteristic of the 2$^{nd}$ layer above graphite. However, the gap amplitude (~ 250 meV) is larger than that in Fig. 2b (~170 meV), as is the thickness (1.1 nm *vs.* 0.95 nm). The 3$^{rd}$ layer exhibits a U-shaped Mott-gap far away from the step (interior of the terrace), while the *dI/dV* spectra reveal a crossover from the large U-shaped Mott-gap to the reduced-gap as the tip approaches the step-edge (Fig. 5h). The crossover process is dominated by the elevation of occupied states and filling holes in the Mott-gap, demonstrating the tight relationship to the in-gap states in the 1$^{st}$ layer. With the emergence of a subband near 300 meV, the UHB transfers its spectral weight to the low energy side. When we compared the thicknesses of the 3$^{rd}$ α-RuCl$_3$ layers in different flaks in Figs. 5b and g, we discovered that the thickness in Fig. 5g (~1.9 nm) is also slightly larger than that in Fig. 5b (~1.75 nm), indicating that the thickness, *i.e.* the interface coupling, is more important than the layer number for the possible Mott transition. The above results show that the possible Mott transition is caused by charge doping from the heterointerface rather than being a layer-dependent phenomenon. The possible Mott transition typically happens in the 2$^{nd}$ and 3$^{rd}$ layers, where the charges from the heterointerface can be transferred. APRES [8] and optoelectronic measurement [39] previously discovered similar Mott transitions in charge doped α-RuCl$_3$, where the Mott-gap is similarly suddenly lowered upon ionic doping. It was thought to be caused by the relationship between doping-charges and significant spin fluctuation [8]. We still don't fully comprehend the possible unconventional Mott-transition in this work. The mechanism of the possible unconventional Mott transition merits more investigation.

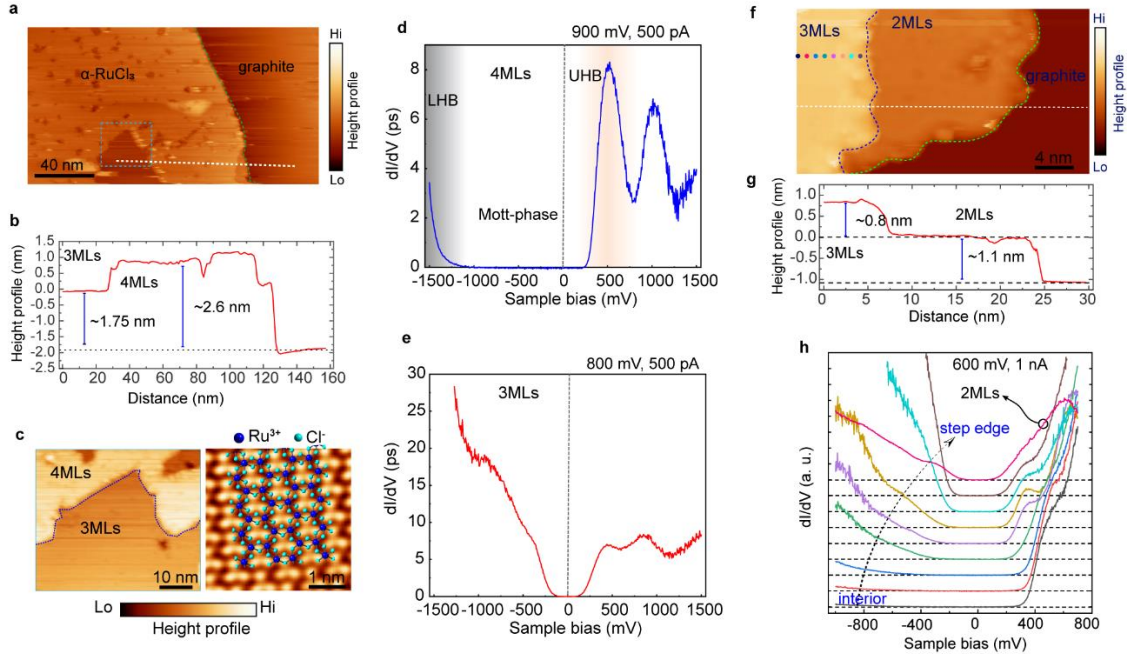

**Figure 5. a**, A broad-view STM image of α-RuCl$_3$ terraces on graphite ($V_H$ =1 V; $I_s$=500 pA); **b**, Height profile along the dashed white line shows the thicknesses of the α-RuCl$_3$ terraces in the squared region correspond to surfaces of the 3$^{rd}$ and 4$^{th}$ layers, respectively; **c**, Left: an enlarge STM image of the 3$^{rd}$ and 4$^{th}$ α-RuCl$_3$ terraces ($V_H$ =1 V; $I_s$=500 pA); right: atomic-resolved STM topography on the 3$^{rd}$ layer filtered by reversed Fourier transform, reveals the sites of Cl atoms; **d** and **e**, Averaged $dI/dV$ spectra acquired on the 4$^{th}$ ($V_H$=900 mV, $I_s$=500 pA) and 3$^{rd}$ ($V_H$=800 mV, $I_s$=500 pA) layers, respectively. Two representative spectra show the possible Mott-phase transition between the 4$^{th}$ and 3$^{rd}$ layers; **f**, STM morphology shows another α-RuCl$_3$ flake on graphite ($V_H$ =600 mV; $I_s$=1 nA); **g**, Height profile along the dashed white line in (f) shows the thicknesses correspond to the surfaces of the 3$^{rd}$ and 2$^{nd}$ α-RuCl$_3$ layers on graphite; **h**, $dI/dV$ spectra are collected precisely at each colored point as marked on the 3$^{rd}$ layer in (f) near the step-edge, with an averaged $dI/dV$ spectrum taken on the 2$^{nd}$ layer for comparison ($V_H$=600 mV, $I_s$=1 nA).

In conclusion, distinct states were detected in the Mott-gap in α-RuCl$_3$ single-layer that is directly contacted to graphite; huge modulations of the tunneling spectra with transferring Hubbard bands to low-energy sides and a dramatically reduced Mott-gap were observed on the upper-layer α-RuCl$_3$ (generally on surfaces of the 2$^{nd}$ and 3$^{rd}$ layers); and the crossover from the large-gapped Mott insulator to the new insulator with reduced-gap was observed at the step-edge in a α-RuCl$_3$ flake. Based on the experimental results, we propose that the overlap of EWFs between graphite and α-RuCl$_3$ at the heterointerface causes in-gap states in the 1$^{st}$ α-RuCl$_3$ layer, and then the charges from the heterointerface are transferred into the upper-layers to induce a weak doping and a possible unconventional Mott transition of α-RuCl$_3$. The findings show an intriguing electronic phenomenon in the doped α-RuCl$_3$ Kitaev QSL system, which warrants future theoretical and experimental exploration.

**Corresponding Author**

*Correspondence to: Xiaohu Zheng, Email: zhengxh@baqis.ac.cn

**Author Contributions**

X. H. Z. and J. H. R. fabricated heterostructure samples. X. H. Z. conducted the STM experiments. K. J., C. L. Y. and Y. G. S. provided the α-RuCl$_3$ samples. X. H. Z. analyzed the data. X. H. Z., X. J. W., K. T., and R. -R. D. wrote the paper. X. H. Z. and R. -R. D. conceived the project. All authors contributed to discussions of the results.

**Acknowledgments.** We would like to thank Deepak Karki for helpful discussion. This work was supported by National Basic Research & Development plan of China (Grants No. 2019YFA0308400), the National Natural Science Foundation of China (Grants No. U2032204), and the Strategic Priority Research Program of the Chinese Academy of Sciences (Grants No. XDB28000000 and XDB33030000).